# Effects of Confinement and Orientation on the Thermoelectric Power Factor of Silicon Nanowires

Neophytos Neophytou and Hans Kosina

Institute for Microelectronics, TU Wien, Gußhausstraße 27-29/E360, A-1040 Wien, Austria

e-mail: {neophytou|kosina}@iue.tuwien.ac.at

## Abstract

It is suggested that low dimensionality can improve the thermoelectric (TE) power factor of a device, offering an enhancement of the *ZT* figure of merit. In this work the atomistic $sp^3d^5s^*$-spin-orbit-coupled tight-binding model and the linearized Boltzmann transport theory is applied to calculate the room temperature electrical conductivity, Seebeck coefficient, and power factor of narrow 1D silicon nanowires (NWs). We present a comprehensive analysis of the thermoelectric coefficients of n-type and p-type NWs of diameters from 12nm down to 3nm, in [100], [110], and [111] transport orientations at different carrier concentrations. We find that the length scale at which the influence of confinement on the power factor can be observed is at diameters below 7nm. We show that contrary to the current view, the effect of confinement and geometry on the power factor originates mostly from changes in the conductivity which is strongly affected, rather than the Seebeck coefficient. In general, enhanced scattering at these diameter scales strongly degrades the conductivity and power factor of the device. We identify cases, however, for which confinement largely improves the channel's conductivity, resulting in ~2-3X power factor improvements. Our results may provide guidance in the design of efficient low dimensional thermoelectric devices.

# I. Introduction

The ability of a material to convert heat into electricity is measured by the dimensionless figure of merit $ZT=\sigma S^2 T/(k_e+k_l)$, where $\sigma$ is the electrical conductivity, $S$ is the Seebeck coefficient, and $k_e$ and $k_l$ are the electronic and lattice part of the thermal conductivity, respectively. Thermoelectric devices have traditionally found use only in niche applications for reasons of low efficiency and high prices. The parameters that determine $ZT$ are interdependent such that $ZT$ remains low. In addition, some of the best thermoelectric materials are rare earth, or toxic materials. Recently, however, low-dimensional thermoelectric devices based on 1D nanowires (NWs), thin films, as well as 1D/2D superlattices [1, 2, 3, 4, 5, 6] and commonly available materials such as Si and Ge have been realized. Low-dimensional materials offer the capability of improved thermoelectric performance. The length scale offers a degree of freedom in engineering $S$, $\sigma$, and $k_l$ through partial control over the dispersions and scattering mechanisms of both electrons and phonons. As a result, enhanced $ZT$ values in nanostructures compared to their bulk material's values were achieved.

Enhanced thermoelectric performance was recently demonstrated for silicon NWs [1, 2]. Although bulk silicon has a $ZT_{bulk} \sim 0.01$, the room temperature $ZT$ of silicon NWs was experimentally demonstrated to be $ZT\sim 0.5$. Most of this improvement has been a result of suppressed phonon conduction ($k_l$) from enhanced boundary scattering [1, 7]. It has also been suggested, however, that low dimensionality can be beneficial for increasing the power factor ($\sigma S^2$) of the device as well [8, 9, 10, 11]. Hicks and Dresselhaus pointed out that channels of lower dimensionality can potentially improve the Seebeck coefficient [8]. The sharp features in the low-dimensional density of states $DOS(E)$ function can improve $S$, as this quantity is proportional to the energy derivative of $DOS(E)$. Mahan and Sofo have further shown that thermoelectric energy conversion through a single energy level (zero-dimensional channel) can reach the Carnot efficiency when the lattice part of the thermal conductivity is zero [12].

Other theoretical studies, focusing on the effect of bandstructure and utilizing ballistic transport, have indeed verified benefits to the power factor from dimensionality [11, 13, 14, 15]. The magnitude of these benefits, however, was only modest <2X. Vo *et al.* in Ref. [16] have consider ab-initio calculations and linearized Boltzmann transport theory under the constant relaxation time approximation to investigate the thermoelectric properties of Si NWs of 1.1nm and 3nm in diameter. Although these diameters are possibly too small for any practical considerations, they have shown that *ZT* values around unity can be achieved if the thermal conductivity is reduced below $k_l$=2W/mK.

In this work, we calculate the thermoelectric coefficients ($\sigma$, *S* and $k_e$) of Si NWs of larger diameters up to *D*=12nm. We employ atomistic calculations for electronic structures and linearized Boltzmann theory [12, 17] for transport. We use the full energy dependence for the relaxation times of the scattering mechanisms considered. We present a comprehensive analysis of the thermoelectric coefficients of cylindrical NWs in terms of: i) n-type and p-type, ii) of diameters from *D*=3nm up to *D*=12nm, iii) in [100], [110] and [111] transport orientations, and iv) for different carrier concentrations. Using experimentally measured lattice thermal conductivity values we estimate the NWs' *ZT*. The focus of this work is on the effects of geometry on the thermoelectric power factor through electronic structure variations. Therefore most calculations are performed at room temperature, but we also show that the main conclusions are valid for other temperatures as well.

We show that at room temperature, *S* can be improved for NWs with diameters below ~7nm [11]. At these diameter scales, however, the scattering mechanisms, and especially surface roughness scattering (SRS), become increasingly important and significantly reduce the power factor. This is the case for all n-type NWs and [100] p-type NWs. Interestingly, however, in p-type [111] and [110] NWs, diameter scaling improves the channels' conductivity [18, 19] and results in enhanced power factors. Our results indicate that the quantity that controls improvements and variations in the power factor is the electrical conductivity, whereas the changes in the Seebeck coefficient as a result of diameter scaling or transport orientation are small, contrary to current thinking, and contrary to results deduced from ballistic calculations [11, 15].

The paper is organized as follows: In section II, we describe the numerical approach: i) the tight-binding model, and ii) the derivation of the scattering rates from linearized Boltzmann theory and the atomistic dispersions. In section III we perform an analysis of the thermoelectric coefficients of n-type NWs and in section IV of p-type NWs. In section V we discuss the effect of temperature, calculate an estimate for the *ZT* figure of merit and discuss the possible implications of NW surface reconstruction on our results. Finally, in VI we conclude.

## II. Approach

### A. **Atomistic modeling:**

To obtain the bandstructure of the NWs both for electrons and holes for which spin-orbit coupling is important, a well calibrated atomistic model is used. The nearest neighbor $sp^3d^5s^*$-SO tight-binding (TB) model [18, 20, 21, 22] captures all the necessary band features, and in addition, is robust enough to computationally handle larger NW cross sections as compared to ab-initio methods. As an indication, the unit cells of the NWs considered in this study contain from ~150 to ~5500 atoms, and the computation time needed varies from a few hours to a few days for each case on a single CPU. Each atom in the NW unit cell is described by 20 orbitals including spin-orbit-coupling. The model itself and the parameterization used [20], have been extensively calibrated to various experimental data of various natures with excellent agreement, without any material parameter adjustments [23, 24, 25, 26, 27]. The model provides a simple but effective way for treatment of the surface truncation by hydrogen passivation of the dangling bonds on the surfaces [28]. What is important for this work, the Hamiltonian is built on the diamond lattice of silicon, and the effect of different orientations is automatically included, which impacts the interaction and mixing of various bulk bands.

We consider cylindrical silicon NWs in three different transport orientations [100], [110], and [111] and diameters varying from $D$=12nm down to $D$=3nm. Strain fields [29, 30], as well as relaxation of the NW surfaces are neglected in this study, but in section V we briefly discuss the possible implications of surface relaxation on our results.

The strong influence of bandstructure in the transport properties of narrow NW and ultra-thin-body channels is stressed in our previous works. For n-type NWs, we previously showed that the effective masses change as the diameter is scaled below 7nm, even up to 90% depending on the orientation [18]. More importantly, for p-type NWs we showed that the carrier velocities can change by a factor of ~2X with confinement (~4X variations in $m^*$), again depending on the orientation. In this case, the subband curvature is in addition sensitive to electrostatic potential variations in the cross section of the NWs, an effect that cannot be captured with simplified effective mass electronic structure models [19, 31, 32]. This behavior is attributed to the strong anisotropy of the heavy-hole band. Indeed, the origin of most of these features can be traced back to and understood from the bandstructure of bulk Si. These dispersion variations will affect the transport properties of the NWs, but can also provide additional degrees of freedom for performance optimization. An electronic structure model capable of capturing such effects is therefore essential in investigating the influence of confinement in ultra narrow NW channels.

### B. Boltzmann theory:

In Linearized Boltzmann formalism, the electrical conductivity ($\sigma$), the Seebeck coefficient ($S$), and the electronic part of the thermal conductivity ($k_e$) are defined as:

$$\sigma = q_0^2 \int_{E_0}^{\infty} dE \left( -\frac{\partial f_0}{\partial E} \right) \Xi(E), \tag{1a}$$

$$S = \frac{q_0 k_B}{\sigma} \int_{E_0}^{\infty} dE \left( -\frac{\partial f_0}{\partial E} \right) \Xi(E) \left( \frac{E - E_F}{k_B T} \right), \tag{1b}$$

$$\kappa_0 = k_B^2 T \int_{E_0}^{\infty} dE \left(-\frac{\partial f_0}{\partial E}\right) \Xi(E) \left(\frac{E-E_F}{k_B T}\right)^2, \tag{1c}$$

$$\kappa_e = \kappa_0 - T\sigma S^2. \tag{1d}$$

The transport distribution function $\Xi(E)$ is defined as [12, 17]:

$$\Xi(E) = \frac{1}{A} \sum_{k_x, n} v_n^2(k_x) \tau_n(k_x) \delta(E - E_n(k_x))$$
$$= \frac{1}{A} \sum_n v_n^2(E) \tau_n(E) g_{1D}^n(E). \tag{2}$$

where $v_n(E) = \frac{1}{\hbar} \frac{\partial E_n}{\partial k_x}$ is the bandstructure velocity, $\tau_n(k_x)$ is the momentum relaxation time for a state in a specific $k_x$-point and subband $n$, and:

$$g_{1D}^n(E_n) = \frac{1}{2\pi\hbar} \frac{1}{v_n(E)} \tag{3}$$

is the density of states for 1D subbands (per spin).

The focus of this work is on the effects of confinement and orientation on the thermoelectric power factor through the modifications of the electronic structure. Since the focus is geometry, all calculations are performed at room temperature and the effect of phonon drag in the calculation of the Seebeck coefficient is not considered in Eq. (1b). Phonon drag is primarily a low temperature effect with temperature dependence $T^{-9/2}$ [33, 34] and has a significant contribution to the thermopower at $T<50K$. At room temperature its contribution is insignificant as discussed in Refs [33, 34, 35, 36]. In section V we discuss the role of temperature, and show that the trends and conclusions we describe at room temperature still hold at different temperatures as well.

### C. Scattering rate calculation:

The transition rate $S_{n,m}(k_x, k_x')$ for a carrier in an initial state $k_x$ in subband $n$ to a final state $k_x'$ in subband $m$ is extracted from the atomistic dispersions and wave form overlaps using Fermi's Golden Rule [37, 38] as:

$$S_{n,m}(k_x, k_x') = \frac{2\pi}{\hbar} \left| H_{k_x',k_x}^{m,n} \right|^2 \delta\left(E_m(k_x') - E_n(k_x) - \Delta E\right). \tag{4}$$

The calculation of the relaxation times involves an integral equation [39, 40, 41, 42]:

$$\frac{1}{\tau_n(k_x)} = \sum_{m,k_x'} S_{n,m}(k_x, k_x') \left(1 - \frac{v_m(k_x')\tau_m(k_x')f_m(k_x')}{v_n(k_x)\tau_n(k_x)f_n(k_x)}\right) \tag{5}$$

While self-consistent solutions of this may be found, this is computationally very expensive, especially for atomistic calculations. Therefore, it is common practice to simplify the problem. For isotropic scattering (ADP, ODP, IVS) the term in the parenthesis reduces to unity upon integration over the solid angle due to symmetry considerations ($S_{n,m}$ is angle independent). In 1D the angle $\vartheta$ can take only two values: $\vartheta = 0$ and $\vartheta = \pi$. The momentum relaxation rates are, therefore, equal to the scattering rates. This holds for any bandstructure. For elastic intravalley, intrasubband scattering (even if anisotropic, i.e. SRS, or impurity scattering), and under the parabolic band assumption, the term simplifies to:

$$1 - \frac{v_m(k_x')\tau_m(k_x')f_m(k_x')}{v_n(k_x)\tau_n(k_x)f_n(k_x)} = 1 - \frac{v_m(k_x')}{v_n(k_x)} = 1 - \frac{\hbar v_m(k_x')/m_m^*}{\hbar v_n(k_x)/m_n^*} = 1 - \frac{|p_x'|}{|p_x|}\cos\vartheta \tag{6}$$

which is the usual term used in the calculation of the momentum relaxation times [38, 39]. Although this is strictly valid for intrasubband transitions only, it is often used for intersubband transitions as well (assuming a weak $k_x$ dependence of $\tau(k_x)$ [41]). In general, however, one should solve the full integral equation. As mentioned by Fischetti *et* al. in Refs. [40, 41], however, often sufficiently accurate results are obtained using the above approximations, without the need to evaluate numerically demanding integral equations. In this work, we calculate the relaxation times by:

$$\frac{1}{\tau_n(k_x)} = \sum_{m,k_x'} S_{n,m}(k_x, k_x') \left(1 - \frac{v_m(k_x')}{v_n(k_x)}\right) \tag{7}$$

The Fermi functions in Eq. (5) cancel for elastic processes. For inelastic, isotropic processes (as all the inelastic processes we consider) the term in brackets reduces to unity anyway after integration over $k_x$. Although admittedly in 1D $\tau(k_x)$ in the numerator and denominator of Eq. (5) can vary with $k_x$, we still drop it as it is commonly done in the literature [43, 44, 45, 46]. This will only affect the SRS and impurity scattering results at larger diameters where intersubband transitions can be important, but at larger diameters the electronic structure approaches bulk like, and the variation of $\tau(k_x)$ with $k_x$ is smaller. Indeed, our mobility calculations at larger diameters approach bulk behavior for all the scattering mechanisms described.

The matrix element between a carrier in an initial state $k_x$ in subband $n$, and a carrier in a final state $k_x'$ in subband $m$, is computed using the scattering potential $U_S(\vec{r})$ as:

$$H_{k_x',k_x}^{m,n} = \frac{1}{\Omega} \int_{-\infty}^{\infty} F_m^*(\vec{R}) e^{-ik_x'x} U_S(\vec{r}) F_n(\vec{R}) e^{ik_x x} d^2R dx, \tag{8}$$

where the total wavefunction of a state is decomposed into a plane wave $e^{ik_x x}$ in the $x$-direction, and a bound state $F_v(\vec{R})$ in the transverse plane $\vec{R}$ and $\Omega$ is the normalization volume.

### *1. Phonon scattering*:

In the case of phonon scattering, we extend the approach described in Ref. [38] for bulk and 2D carriers, to 1D carriers. The perturbing potential is defined as:

$$U_S(\vec{r}) = A_{\vec{q}} K_{\vec{q}} e^{\pm i(\vec{q}\cdot\vec{r} - \omega t)} \tag{9}$$

where $A_{\vec{q}}$ is associated with the lattice vibration amplitude, $K_{\vec{q}}$ with the deformation potential, and momentum transfer $\vec{q}\cdot\vec{r} = \vec{Q}\cdot\vec{R} + q_x x$. In this case, the matrix element is:

$$H_{k_x',k_x}^{m,n} = \int_{-\infty}^{\infty} \frac{F_m^*(\vec{R})}{\sqrt{A}} \frac{e^{-ik_x'x}}{\sqrt{L_x}} A_{\vec{q}} K_{\vec{q}} e^{\pm i\vec{Q}\cdot\vec{R}} e^{\pm iq_x x} \frac{F_n(\vec{R})}{\sqrt{A}} \frac{e^{ik_x x}}{\sqrt{L_x}} d^2R dx, \tag{10}$$

where $L_x$ is the length of the unit cell. The integral over the transport direction $x$ becomes a Kronecker-delta expressing momentum conservation in the transport direction,

$$H_{k_x',k_x}^{m,n} = I_{k_x',k_x}^{m,n}(\vec{Q}) A_{\vec{q}} K_{\vec{q}} \delta_{k_x',k_x \pm q_x} \tag{11}$$

with

$$I_{k_x',k_x}^{m,n}(\vec{Q}) = \frac{1}{A}\int_R \rho_{k_x',k_x}^{m,n}(\vec{R}) e^{\pm i\vec{Q}\cdot\vec{R}} d^2R, \tag{12a}$$

$$\rho_{k_x',k_x}^{m,n}(\vec{R}) = F_{m,k_x'}(\vec{R})^* F_{n,k_x}(\vec{R}). \tag{12b}$$

When integrating and taking the square of the matrix element, the integral for the form factor is also evaluated as:

$$\left|I_{k_x',k_x}^{m,n}\right|^2 = \frac{1}{A^2}\int_R d^2R \int_{R'} d^2R' \, \rho_{k_x',k_x}^{m,n}(\vec{R})^* \rho_{k_x',k_x}^{m,n}(\vec{R}) e^{\pm i\vec{Q}\cdot(\vec{R}-\vec{R}')} \tag{13}$$

The summation over the lateral momentum before substitution into Eq. (7) can be performed as:

$$\sum_{k_{\vec{R}}} \left|I_{k_x',k_x}^{m,n}\right|^2 = \frac{A}{4\pi^2}\frac{1}{A^2}\int_Q d^2Q\, e^{\pm i\vec{Q}\cdot(\vec{R}-\vec{R}')}\int_R d^2R \int_{R'} d^2R' \, \rho_{k_x',k_x}^{m,n}(\vec{R})^* \rho_{k_x',k_x}^{m,n}(\vec{R})$$

$$= A\int_R \left|\frac{\rho_{k_x',k_x}^{m,n}(\vec{R})}{A}\right|^2 d^2R \tag{14}$$

$$= \frac{A}{A_{k_x',k_x}^{m,n}}$$

Here, $1/A_{k_x',k_x}^{m,n}$ has the units of m$^{-2}$. For computational efficiency of the calculation of the form factor overlap, on each atom we add the probability density of the components of each multi-orbital wavefunction, and afterwards perform the final/initial state overlap multiplication. In such way, we approximate the form factor components of a lattice atom at a specific location $R_0$ by:

$$\left|\rho_{k_x',k_x}^{m,n}\right|^2 = \sum_\alpha F_{n,k_x}^\alpha F_{m,k_x'}^{\alpha\,*} \sum_\alpha F_{n,k_x}^\alpha F_{m,k_x'}^{\alpha\,*}$$

$$\approx \sum_\alpha F_{n,k_x}^\alpha F_{n,k_x}^{\alpha\,*} \sum_\alpha F_{m,k_x'}^\alpha F_{m,k_x'}^{\alpha\,*} = \sum_\alpha \left|F_{n,k_x}^\alpha\right|^2 \sum_\alpha \left|F_{m,k_x'}^\alpha\right|^2 \equiv \left\|F_{n,k_x}\right\|^2 \left\|F_{m,k_x'}\right\|^2 \tag{15}$$

where $\alpha$ runs over the TB orbitals of a specific atom. This computes the overlaps using the probability density of each state, as in a single orbital (i.e. effective mass) model, although we still keep the $k_x$-dependence of the wavefunctions. Indeed, our numerical overlaps agree with the analytical expressions for the wavefunction overlaps if one assumes sine/cosine wavefunctions and parabolic bands, which can be derived to be $\frac{1}{A_{nm}} = \frac{9}{4A}$ for intra-band and $\frac{1}{A_{nm}} = \frac{1}{A}$ for inter-band transitions [13, 38].

The approximation in Eq. (15) is important because it reduces the memory needed in the computation by 20X, allowing simulations of large NW cross sections with only minimal expense in accuracy. Even after this simplification, the storage of the probability density for the larger diameter NWs requires several Giga bytes of memory. It would have been computationally prohibitive to perform the calculations for the larger diameter NWs using the actual wavefunctions, at least on a single CPU.

The transition rate for phonon scattering is then given by:

$$S_{n,m}(k_x, k_x') = \frac{2\pi}{\hbar} \left| I_{k_x',k_x}^{m,n} \right|^2 \left| K_{\vec{q}} \right|^2 \left| A_{\vec{q}} \right|^2 \delta_{k_x',k_x \pm q_x} \delta\left(E_m(k_x') - E_n(k_x) \pm \hbar\omega_{ph}\right) \quad (16)$$

where $\left| A_{\vec{q}} \right|^2 = \frac{1}{\Omega} \frac{\hbar\left(N_\omega + \frac{1}{2} \mp \frac{1}{2}\right)}{2\rho\omega_{ph}}$, $\rho$ is the mass density, and $N_\omega$ is the number of phonons given by the Bose-Einstein distribution.

The relaxation rate of a carrier in a specific subband $n$ as a function of energy is then given by:

$$\frac{1}{\tau_{ph}^n(E)} = \frac{\pi}{\hbar} \frac{\left(N_\omega + \frac{1}{2} \mp \frac{1}{2}\right)}{\rho\hbar\omega_{ph}} \\ \times \left( \frac{1}{L_x} \sum_m \frac{\left|K_{\vec{q}}\right|^2}{A_{nm}^{k_x k_x'}} \delta_{k_x',k_x \pm q_x} \delta\left(E_m(k_x') - E_n(k_x) \pm \hbar\omega_{ph}\right)\left(1 - \frac{v_m(k_x')}{v_n(k_x)}\right)\right), \quad (17)$$

where $\hbar\omega_{ph}$ is the phonon energy, and we have used $\Omega = AL_x$. For acoustic phonon scattering (ADP or IVS), it holds that $\left|K_{\vec{q}}\right|^2 = q^2 D_{ADP}^2$, whereas for optical phonon scattering (ODP for holes, IVS for electrons) it holds that $|K_{\vec{q}}|^2 = D_O^2$, where $D_{ADP}$ and $D_0$ are the scattering deformation potential amplitudes. For inter-valley scattering (IVS) in the conduction band we include all relevant *g*- and *f*-processes. Specifically for elastic acoustic deformation potential scattering (ADP), after applying the equipartition approximation, the relaxation rate becomes:

$$\frac{1}{\tau_{ADP}^n(E)} = \frac{2\pi}{\hbar} \frac{D_{ADP}^2 k_B T}{\rho v_s^2} \left( \frac{1}{L_x} \sum_{m,k_x'} \frac{1}{A_{nm}^{k_x k_x'}} \delta_{k_x',k_x \pm q_x} \delta(E_m(k_x') - E_n(k_x)) \left(1 - \frac{v_m(k_x')}{v_n(k_x)}\right) \right), \quad (18)$$

where $v_s$ is the sound velocity in Si.

### *2. Surface roughness scattering*:

For SRS, we assume a 1D exponential autocorrelation function [47] for the roughness given by:

$$\langle \delta(x) \delta(x') \rangle = \Delta_{rms}^2 e^{-\sqrt{2}|x-x'|/L_C} \quad (19)$$

with $\Delta_{rms} = 0.48$nm and $L_C = 1.3$nm [46]. Surface roughness is assumed to cause a band edge shift. The scattering strength is derived from the shift in the band edges with quantization $\frac{\Delta E_{C,V}}{\Delta D}$ [48, 49]. The transition rate is derived as:

$$S_{n,m}^{SRS}(k_x, k_x') = \frac{2\pi}{\hbar} \left( q_0 \frac{\Delta E_{C,V}}{\Delta D} \right)^2 \left( \frac{2\sqrt{2} \Delta_{rms}^2 L_C}{2 + q_x^2 L_C^2} \right) \delta(E_m(k_x') - E_n(k_x)), \quad (20)$$

where $q_x = k_x - k_x'$. This surface roughness model is more simplified than the ones described Refs. [46, 50, 51, 52], that account for additional Coulomb effects, the wavefunction deformation at the interface, and the position of electrons in the channel. In this work, however, we ignore these effects since they only cause quantitative changes in our results. We assume that they are lumped into an enhanced roughness $\Delta_{rms}$. Qualitative trends in this work mostly originate from geometry-induced electronic structure variations. In addition, we only consider channels with flat potential in their cross section, in which case the probability density is mostly concentrated in the middle of the channel and is relatively flat near the interface [31], which can partially justify our approximation. As described by various authors, the band edge variation is the dominant SRS mechanism in ultra scaled channels [46, 48, 49, 53, 54]. In Refs [46, 49] it was shown that that the SRS limited low-field mobility in ultra thin nanostructures follows a $L^6$ behavior, where $L$ is the confinement length scale, originating from subband shift due to quantization.

### *3. Impurity scattering:*

Assuming the $x$ direction to extend to infinity, the scattering potential for screened Coulomb ionized impurities is given by:

$$U_S(\vec{r}) = \frac{q_0^2}{4\pi\kappa_s\varepsilon_0} \frac{e^{-\sqrt{(\vec{R}-\vec{R}')^2 + x^2}/L_D}}{\sqrt{(\vec{R}-\vec{R}')^2 + x^2}}, \qquad (21)$$

where $\vec{R}$ is the position of an electron in the 2D cross section at $x=0$, feeling the influence of an impurity at $(x, \vec{R}')$. The screening length $L_D$ is given by:

$$L_D = \sqrt{\frac{\kappa_s\varepsilon_0 k_B T}{q_0^2 n} \frac{\Im_{-1/2}(\eta_F)}{\Im_{-3/2}(\eta_F)}}. \qquad (22)$$

where $\Im_\alpha(\eta_F)$ is the Fermi-Dirac integral of order $\alpha$ and $n$ is the carrier concentration. The matrix element for an impurity-electron scattering then becomes:

$$H_{k_x',k_x}^{m,n}(\vec{R}') = \int_R \frac{\|F_{n,k_x}(\vec{R})\|}{\sqrt{A}} \left( \frac{1}{L_x} \int_{-\infty}^{\infty} \frac{q_0^2 e^{iq_x x}}{4\pi\kappa_s\varepsilon_0} \frac{e^{-\sqrt{(\vec{R}-\vec{R}')^2 + x^2}/L_D}}{\sqrt{(\vec{R}-\vec{R}')^2 + x^2}} dx \right) \frac{\|F_{m,k_x'}(\vec{R})\|}{\sqrt{A}} d^2R \qquad (23)$$

where the expression in the parenthesis is the Green's function of the infinite channel device. For a cylindrical channel, the expression in the parenthesis is the modified Bessel function of second kind of order zero, $K_0(q, \vec{R}')$ [46, 55, 56, 57]. Again, as in the case of phonon scattering, we have used the probability density on each atomic site instead of the actual wavefunctions, which largely reduces the computational complexity.

The total transition rate due to impurity scattering is computed after taking the square of the matrix element, multiplying by the number of impurities in the normalized cross sectional area of the NW in the length of the unit cell ($N_I L_x$), and integrating over the distribution of impurities in the cross sectional area of the NW (over $\vec{R}'$). The impurities are assumed to be uniformly distributed on the atomic lattice points. The transition rate is then given by:

$$S_{n,m}^{imp.}(k_x,k_x') = \frac{2\pi}{\hbar}\left(\int_{R'}(N_I L_x)\left|H_{k_x',k_x}^{m,n}(\vec{R}')\right|^2 d^2R'\right)\delta\left(E_m(k_x')-E_n(k_x)\right) \quad (24)$$

### D. General scattering considerations:

#### *1. Scattering state selection*:

Elastic and inelastic scattering processes are considered. We consider bulk phonons, and *bulk* Si scattering "selection rules". For n-type NWs, the elastic processes (elastic phonons, SRS, impurity scattering) are only treated as intra-valley, whereas inelastic ones (inelastic phonons) are only treated as inter-valley (IVS). Although all valleys in the electronic structure of bulk Si collapse from 3D to 1D k-space in our calculations, we carefully chose the final scattering states for each event such that we follow the *bulk* processes. For inelastic transitions all six *f*- and *g*-type processes in Si are included [38, 58]. For p-type, we consider ADP and ODP processes which can be intra-band and inter-band as well as intra-valley and inter-valley.

#### *2. Bulk vs. confined phonons*:

The full dispersion of confined phonons is neglected. Bulk phonons provide an ease of modeling and allow the understanding the effects on bandstructure on the thermoelectric coefficients, still with good qualitative accuracy in the results. Spatial confinement mostly affects acoustic phonons by "bending" of the acoustic modes, which results in lower than the bulk "effective" group velocity [59]. Optical phonons are not affected significantly by confinement [60, 61]. Such effects, however, will only affect our results quantitatively (and as we show below, only slightly), whereas the qualitative behavior, which originates from the influence of geometry on bandstructure will not be affected. As described by several authors, the effect of phonon confinement on the mobility for the thinnest NWs examined in this work can be of the order of 10-20% (further reduction in mobility) and declines fast as the diameter increases [43, 44, 45, 60]. Especially in the case of p-type channels, studies have shown that using confined or bulk phonons makes very little difference in mobility calculations [62]. Similar studies in GaAs NWs also indicated the same result [63]. These observations, however, strongly depend on the boundary conditions one uses for the calculation of the confined phonon modes, which introduces an additional uncertainty [43, 44, 60].

The deformation potential parameters we use are the same as in Ref. [38], with the exceptions of $D_{ODP}^{holes} = 13.24 \times 10^{10} \, \text{eV/m}$, $D_{ADP}^{holes} = 5.34 \, \text{eV}$, and $D_{ADP}^{electrons} = 9.5 \, \text{eV}$ from Refs [37, 43, 44] which are more relevant for NWs. These are higher than the bulk values [38]. It is common practice to employ larger electron-phonon deformation potential values to explain phonon-limited transport trends in nanostructures [46, 64, 65, 66], which could partially account for phonon confinement effects. Buin *et* al. [43, 44] calculated the mobility of NWs up to *D*=3nm using the same TB model and both confined and bulk phonons. In this work, by using the same parameters we were able to benchmark our results for mobility for the *D*=3nm NW bulk phonon case with a good agreement, before extending to larger diameters. As discussed in Ref. [43, 44], however, the most accurate deformation potentials for NWs may finally be obtained by comparing mobility with experimental data [43, 64, 65], which for NWs at this time are sparse. Different sets of parameters will indeed change the magnitude of the phonon-limited results, in particular that of the electrical conductivity. The thermoelectric coefficient trends with diameter and orientation, however, will not be affected.

## *2. Strength of each scattering mechanism*:

Clearly, the presence of additional scattering mechanisms (interface roughness scattering and especially impurity scattering at high impurity concentrations) also limits the conductivity, and the relative effect of phonons and especially acoustic phonon confinement on the total conductivity will be even smaller [60]. This is illustrated in Table I, which reports the percentage contributions to the total scattering rate of the four different scattering mechanisms separately (column-wise): ADP (long wavelength intra-valley acoustic), IVS (inter-valley, inelastic), SRS and impurity scattering. The results are for the narrowest $D$=3nm n-type, [100] NW, for which the effect of acoustic phonon confinement will be the strongest. The rows show cases of different scattering situations. The individual contributions are extracted by weighting each energy dependent relaxation rate by the Fermi distribution and integrate over all energy (carrier concentration $n=10^{18}$/cm$^3$ is assumed for placing the Fermi level for rows one-five and $n=10^{19}$/cm$^3$ for rows six-seven). Phonon scattering is always included, whereas SRS and impurity scattering are gradually added. The first row shows that if only phonon scattering is considered, the effect of ADP and IVS are almost equal. When SRS is included (second and third rows), most of the contribution to the relaxation rate originates from SRS. For roughness $\Delta_{rms}^{1}$=0.24nm (second row) the contribution of the ADP scattering to the total scattering rate is 17% and drops to 6% when the $\Delta_{rms}$ increases to $\Delta_{rms}^{2}$=0.48nm (the one used throughout the manuscript). Once impurity scattering is considered (assuming $n_0=10^{18}$/cm$^3$) in the fourth and fifth row under weak and stronger SRS respectively, then the contribution of ADP drops to 6% and 4%. In rows six and seven we increase the impurity concentration to $n_0=10^{19}$/cm$^3$ (the one used throughout the manuscript). The ADP contribution drops even further to only 4% and 3% for the cases of weak and stronger SRS. Impurity scattering at such high concentrations dominates the scattering processes. Indeed, this is in agreement with results based on impurity scattering in bulk Si where it is well known that the mobility drops by almost an order of magnitude from the phonon-limited value at such high concentrations [67]. The main point is that SRS and impurity scattering dominate the scattering rates. The contribution of acoustic phonons to the total scattering rate is not large once these two mechanisms are considered. Using bulk phonons for simplicity instead of confined phonons, therefore, might slightly overestimate our conductivity results, but it would not alter the qualitative trends we present and the conclusions of this work.

We note that similar atomistic studies that couple semi-empirical tight-binding (or DFT) to Boltzmann transport are also discussed in the literature for either NWs of smaller diameters, fewer scattering processes, or different channel structures and materials [16, 44, 68, 69]. In this work we focus on the effect of dimensionality and geometry, how it influences the electronic structure, and through this the TE coefficients. We consider relatively "large" diameters up to *D*=12nm in order to identify the transition of the thermoelectric power factor from 3D bulk-like (*D*=12nm) to 1D like behavior (*D*=3nm), since the electronic structure of Si at *D*=12nm is already almost bulk-like. We have made several approximations when computing the scattering rates, but we believe that the effect of these on the results will only be quantitative, since any qualitative behavior is mostly electronic structure related.

## III. n-type NW TE coefficients

In this part, a comprehensive analysis of the thermoelectric coefficients in n-type NWs of different orientations and diameters is performed.

### A. **n-type NWs: Phonon limited TE coefficients**

Since the Seebeck coefficient is proportional to the energy derivative of the *DOS*, it has been suggested that low dimensional channels, with sharp features in their energy *DOS(E)* function, would be beneficial for thermoelectric devices [8, 9]. On the other hand, the electrical conductivity degrades in nanostructures due to stronger phonon scattering and enhanced SRS. The two processes act inversely proportional in defining the power factor $\sigma S^2$. A careful analysis of the two quantities is needed to determine whether benefits to the power factor $\sigma S^2$ can be expected from ultra narrow low dimensional materials.

Figure 1 shows the thermoelectric coefficients $\sigma$, *S*, and $\sigma S^2$ of n-type NWs in the [100] (blue), [110] (red) and [111] (green) transport orientations, for *D*=3nm (solid) and *D*=12nm (dashed) as a function of the electron concentration. In this case, only phonon scattering is considered. Further on we will also consider the effect of SRS and impurity scattering. Figure 1a shows the conductivity of the NWs. At the same carrier concentration, the conductivity of the NWs with the smaller diameter *D*=3nm, is reduced from that of the *D*=12nm diameter NWs.

Again, comparing at the same carrier distribution, the Seebeck coefficient in Fig. 1b increases as the diameter reduces. For smaller diameters some orientation dependence is observed, due to the different electronic structures of the NWs. This anisotropy is minimized for larger diameters both for the conductivity and especially for the Seebeck coefficient. In Ref. [16], Vo *et* al., have also observed some orientation dependence. The orientation dependence, however, in all cases is small to make a strong case for a preferential design direction for n-type NW thermoelectric devices.

Figure 1c shows the power factor of the n-type NWs. The trends are similar for NWs in all transport orientations. The power factor of all $D$=3nm NWs is reduced compared to the power factor of the $D$=12nm NWs, except for very high carrier concentrations, following the trend of electrical conductivity shown in Fig. 1a. Some orientation dependence can be observed, following the same order and relative magnitude as the conductivity in Fig. 1a. It is evident from this, that the effect of the electrical conductivity is stronger in influencing the power factor than that of the Seebeck coefficient. Therefore, although narrower NWs can be beneficial to the Seebeck coefficient, thinning the diameter has a stronger degrading effect on the conductivity due to enhanced scattering, and the power factor is overall reduced.

The difference in the behavior of the NWs of different diameter and orientation in both $\sigma$ and $S$ originates from the position of the Fermi level with respect to their band edge $E_C - E_F$. The closer $E_C$ resides to $E_F$, the higher is the conductivity and the lower the Seebeck coefficient. The $E_C - E_F$ is plotted in Fig. 2 versus the carrier concentration for all the NWs of Fig. 1. The bandedges of all $D$=12nm NWs reside at almost the same distance away from $E_F$, and therefore a very small anisotropy is observed, especially in $S$. The bandedges of the $D$=3nm NWs reside farther from $E_F$ compared to those of the $D$=12nm NWs. This may appear counter intuitive since one would have expected that the smaller number of subbands in the thinner NW will force $E_C$ to be closer to $E_F$ at the same carrier concentration. It appears this way because the carrier concentration is normalized to the diameter. Considering a NW at a specific carrier concentration and specific $E_C$-$E_F$, as the diameter is reduced, if the number of subbands in the energy region of relevance decreases linearly with the cross sectional area of the NW, then $E_C$-$E_F$ will stay the same. Once a single subband remains, reducing the diameter does not further reduce the number of subbands. To keep the carrier concentration constant with decreasing diameter, $E_C$-$E_F$ increases.

At $D$=3nm, the differences in $\sigma$ and S originate from the different effective masses and degeneracies of the subbands, which control $E_C - E_F$. At a given carrier concentration, the smaller the effective mass and degeneracy are, the closer $E_F$ resides to $E_C$. In such case, the conductivity is higher and the Seebeck coefficient lower. The bandedge of the [110] NW, with a two-fold degenerate lowest valley and the lightest mass ($m^*$=0.16$m_0$) [18] resides closer to $E_F$ compared to the other two NW types. The [100] NW with four-fold degenerate valley of mass $m^*$=0.27$m_0$ follows, whereas the [111] NW, with six-fold degenerate valleys and heavier mass ($m^*$=0.55$m_0$), resides the furthest from $E_F$ compared to the other two NW types. As described in Ref. [18], confinement changes the masses of the NW subbands compared to the bulk $m^*$=0.19$m_0$ in [110], $m^*$=0.19$m_0$ in [100] and $m^*$=0.43$m_0$ in [111] directions [18, 70, 71].

As a result, the conductivity of the [110] NW is the largest, followed by the [100] NW, whereas that of the [111] is the lowest (Fig. 1a). The reverse is observed for the Seebeck coefficient, since the two quantities are inversely related (Fig. 1b). For the $D$=3nm NWs, therefore, some degree of orientation dependence can be observed, which was not evident for the larger diameter NWs. The power factor follows the orientation and magnitude trend of $\sigma$, rather than that of $S$, despite the fact that it depends linearly on $\sigma$ but quadratically on $S$. Changes in $\sigma$ are much larger than changes in $S$, because $\sigma$ depends exponentially on $E_C - E_F$, whereas $S$ depends linearly. At least for carrier concentrations up to $n$=5x10$^{19}$/cm$^3$, $\sigma S^2$ is lower for the $D$=3nm NWs (Fig. 1c) following the trend of $\sigma$. The influence of confinement and geometry, therefore, is stronger on the conductivity which affects the power factor stronger. We note here that the increase in $S$ with diameter reduction is only partially the improvement described by Hicks and Dresselhaus in Refs [8, 9] which originates from the shape of the 1D $DOS(E)$ function.

That effect will be more pronounced when comparing at the same $E_C - E_F$ rather than the same carrier concentration [13]. What we describe can be explained by changes in $E_C - E_F$, which are determined from the normalization of the charge distribution by the cross sectional area and the rate at which the number of subbands in the NW are reduced compared to the reduction of the cross sectional area.

### B. <u>n-type NWs: The effect of confinement</u>

From Fig. 1, the peak of the power factor appears around electron concentrations of $n=10^{19}/cm^3$. In order to clearly observe the diameter dependence on the results, Fig. 3 presents the thermoelectric coefficients of all NWs at $n=10^{19}/cm^3$ versus their diameter. At this carrier concentration, which is high but practically achievable, the Fermi level resides very close to the band edges for most of the NW cases as seen in Fig. 2. The band edge of the $D=3$nm [111] NW is the furthest away from the Fermi level at this concentration, which causes the peak in its power factor to appear at concentrations beyond $10^{20}/cm^3$ (see Fig. 1c), which are too high. The dashed lines in the sub-figures of Fig. 3 indicate the phonon limited thermoelectric coefficients. The solid lines additionally include the effect of SRS. Figure 3a shows the electrical conductivity. Clearly, in all cases the conductivity degrades as the diameter reduces. The phonon limited degradation varies from 0.5X to ~3X depending on the orientation, with the [110] oriented NWs being less affected as also described in Ref. [72]. Once SRS is included, a further ~2X conductivity reduction is observed for the smaller diameter NWs. SRS is weaker for the $D=12$nm NWs. The Seebeck coefficient in Fig. 3b increases as the diameter reduces, a reverse trend compared to conductivity. The increase is at most ~2X, and as explained earlier it originates from fact that $E_C - E_F$ is larger for NWs of smaller diameters. SRS only causes a slight additional increase in $S$, indicating that $S$ is at first-order scattering independent. The power factor in Fig. 3c reduces with diameter, especially when SRS is included in the calculation. The increase in $S$ for lower diameters cannot compensate the large degradation in conductivity. For the cases of [110] and [111] NWs, a maximum can be obtained around $D=7$nm, whereas for the [100] NWs the power factor is reduced monotonically with diameter reduction. From this figure, the [100] NW performs slightly better at *D*=12nm, the [110] at *D*=3nm, whereas in the intermediate diameter ranges, the [111] NWs are slightly advantageous. This orientation dependence is, however, small.

We mention here that ballistic simulations using the Landauer approach [11, 13, 15] suggest the contrary, that diameter scaling will improve the power factor. It is because the ballistic approach captures the improvement in the Seebeck coefficient which is scattering independent at first order, but not the degradation in the conductivity which has a stronger sensitivity to scattering mechanisms and geometry.

### C. <u>n-type NWs: The effect of impurity scattering</u>

Carrier concentrations of $n=10^{19}$/cm$^3$ at which the peak of the power factor appears can be achieved with different methods. Since direct impurity doping is what is traditionally used in TE devices, we demonstrate here the effect of such a high impurity doping on the thermoelectric coefficients. In Fig. 4, we show $\sigma$, $S$, and $\sigma S^2$ as in Fig. 3, but we now include phonons, SRS, and impurity scattering (solid lines). We still show the phonon limited results in dashed lines for comparison.

Figure 4a shows the electrical conductivity. Clearly, in all cases, impurity scattering causes a strong degradation in the conductivity at the entire diameter range. For smaller diameters the degradation is caused by both the SRS and impurity scattering, however at larger diameters the degradation is mostly due to impurity scattering. The degradation is of the order of 6X-8X depending on the orientation and diameter, with the [110] oriented NWs being less affected. This is consistent with the almost one order of magnitude drop in mobility observed for MOSFET devices at such high doping concentrations [67]. The Seebeck coefficient in Fig. 4b increases with the introduction of impurity scattering in the calculation by ~20% if one compares the solid versus dashed lines (phonon limited). The power factor in Fig. 4c decreases due to impurities by ~4X following the conductivity reduction trend, since this quantity is affected more than the Seebeck coefficient.

From these results it is obvious, therefore, that direct doping of the channel will cause large degradation in performance, and alternative strategies to achieve high carrier densities should be employed for high efficiency thermoelectric devices as also mentioned by Ryu *et* al. [66]. Such methods can be: i) gated channels [66, 73], or ii) remote modulation doping (or charge transfer) techniques [74, 75]. Another important observation is that similar to the effect of geometry, introduction of additional scattering mechanisms affect the conductivity much stronger than the Seebeck coefficient. Power factor optimization strategies in low dimensional channels, therefore, should focus in improving (or not reducing) $\sigma$ and not necessarily in optimizing *S*.

## IV. p-type NW TE coefficients

A possible way to improve a channel's conductivity is through bandstructure engineering. Strain is a possible direction which has been applied to MOSFET channel devices with success [29]. Here we describe a different mechanism in p-type channels that originates solemnly from confinement. In certain cases the electronic structure undergoes significant changes with confinement [19]. Through careful engineering of the transport and confinement orientations as well as feature size, the subbands can become lighter and large improvements in conductivity can be achieved [19, 31, 32]. This is shown in Fig. 5 for p-type [110] and [111] NWs. In Fig. 5a and 5b, we show the (100) and (11-2) heavy-hole band bulk Si energy surfaces. The bandstructure of the [110] and [111] NWs, respectively, will be formed by bands of high curvature, residing away from the center of each of these surfaces as shown by the lines in the figures and described in Ref. [19, 31]. The actual NW subband envelopes for diameters ranging from *D*=12nm to *D*=3nm are shown in Fig. 5c and 5d. This increase in the subband curvature with confinement can provide an improvement in the carrier velocities by ~2X [19]. The effective masses of the bands in these NWs can reduce from $m^*$=0.4$m_0$ to values below $m^*$=0.2$m_0$ solely due to confinement [31, 37, 44, 72]. On the other hand, the mass of the [100] NWs still remains high ($m^*$~$m_0$).

In this part, a comprehensive analysis of the thermoelectric coefficients in p-type NWs of different orientations and diameters is performed. We follow the same approach as for n-type NWs above.

## A. p-type NWs: Phonon-limited TE coefficients

Figure 6 shows the electrical conductivity, Seebeck coefficient and power factor for p-type NWs of $D$=3nm (solid lines) and $D$=12nm (dashed lines) in the [100] (blue), [110] (red) and [111] (green) transport orientations versus the hole concentration. Only phonon scattering is considered here. In contrast to what was shown in Fig. 1a for n-type NWs, the electrical conductivity in Fig. 6a is strongly anisotropic and diameter dependent. At $D$=12nm, the [111] NW has the largest conductivity, followed by the [110] NW, whereas the [100] NW lacks behind. As the diameter reduces, the conductivity of the [100] NW reduces, a similar behavior as in n-type NWs. The conductivity of the [111] and [110] NWs, however, strongly increases, originating from the large curvature increase in the NW dispersions with diameter decrease as explained earlier.

For the Seebeck coefficients in Fig. 6b, only that of the [100] p-type NW increases as the diameter scales to $D$=3nm. For the [110] and [111] NWs, $S$ slightly suffers since it is inversely proportional to the electrical conductivity, which undergoes a large increase. Only a small degradation is observed in $S$, because on the one hand $E_C - E_F$ tends to increase with diameter scaling, but on the other hand the reduction in the effective masses of the bands tend to reduce $E_C - E_F$. Finally, $E_C - E_F$ does not change much and $S$ is affected only slightly. The power factor for the [111] and the [110] NWs in Fig. 6c largely increases by ~3X for the smaller diameter NWs, in contrast to the [100] NW power factor which only changes marginally.

As in the case of n-type NWs, the power factor is strongly influenced by the conductivity, rather than the Seebeck trend. Benefits to the power factor can therefore be achieved mainly though bandstructure modifications that affect the conductivity, and less through improvements in the Seebeck coefficient.

### B. p-type NWs: The effect of confinement

Figure 7 shows the p-type NW thermoelectric coefficients versus diameter at $p=10^{19}/cm^3$, approximately the hole concentration at which the power factors peak in Fig. 6c. The dashed lines indicate results for which only phonon scattering is included, whereas the solid lines results for which phonons and SRS are included. Figure 7a shows clearly that the phonon-limited conductivity of [111] and [110] NW orientations is increased by more than a factor of 8X as the diameter is scaled. This increase is large enough to compensate for the effect of SRS. Once SRS is included, the conductivity still remains a factor of ~2X higher for the smaller diameters compared to the larger ones. Strong orientation dependence is evident. The [111] NWs provide ~2X higher conductivity than the [110] ones. The conductivity of the [100] NWs is much lower, and is additionally decreased with diameter scaling.

Strong orientation and diameter dependence is also evident in the Seebeck coefficients of Fig. 7b. At larger diameters, $S$ is almost the same for all NW orientations. As the diameter is reduced, $S$ increases in the [100] NW case, whereas it decreases in the [110] and [111] cases. At the smaller diameters, the [100] NWs have almost ~2X larger $S$ than the other two orientations. The ~2X advantage in $S$ of the [100] NWs over the other orientations is not enough to compensate for the large differences in their conductivity. As a result, the power factor shown in Fig. 7c, is higher for the [111] and [110] NWs, whereas still remains low for the [100] NWs. Including SRS in the results, increases $S$ only slightly. Therefore, the ~2X reduction in conductivity due to SRS, directly translates to a power factor reduction.

## V. Analysis and Discussion

### A. The effect of temperature

Up to this point, our analysis was performed at room temperature, and we have focused on the effect of geometry, confinement, and orientation on the thermoelectric coefficients. Here, we show that the main trends and conclusions we have presented hold at different temperatures as well. The effect of temperature changes the strength of phonon scattering and the shape of the Fermi distribution. The geometry trends, however, are still unchanged. This is shown in Fig. 8 for the n-type [111] NW, but the trends are very similar for the rest of the NW families as well.

Figure 8a shows the phonon-limited electrical conductivity of the NW as a function of diameter for three different temperatures, $T$=150K, 300K and 450K. The carrier concentration is assumed to be $n=10^{19}$/cm$^3$. Phonon scattering strength increases as the temperature increases, and therefore the conductivity decreases with temperature increase at all diameter ranges. The Seebeck coefficient on the other hand, in Fig. 8b, follows the reverse trend, increasing with raising temperature. Part of the reason has to do with the increase in $E_C - E_F$ as the temperature increases. The inset of Fig. 8b shows the $E_C - E_F$ for the $D$=12nm NW as a function of the carrier concentration for the three different temperatures. At higher temperatures, the Fermi distribution spreads out more, filling up more states in the bands. To maintain the same carrier concentration, $E_F$ shifts farther away from the band edge $E_C$, which increases the Seebeck coefficient. The temperature dependence is stronger at larger diameters $D$=12nm, whereas at $D$=3nm it is negligible. The dispersion of the $D$=12nm NW consists of many more subbands. More carriers are picked as the temperature increases, which causes a larger $E_C - E_F$ shift. The power factor in Fig. 8c is more affected by the larger conductivity changes, and therefore reduces with temperature increase. The basic trends are retained for the rest of the NW types as well.

Here, we note that we do not consider the effect of phonon drag in the calculation of the Seebeck coefficient in the analysis above, as we focus on the effect of geometry. Inclusion of phonon drag is significantly demanding involving several phonon scattering mechanisms. Reference [2] points out the importance of phonon drag for $T$~150K, and, therefore, our calculated $S$ at this temperature could be underestimated. In that work, it was claimed that phonon drag can enhance the $ZT$ significantly and further analysis of this concept will be very promising and useful [35]. In general, however, such an effect is not observed for temperatures above $T$~100K, especially for nanostructures where boundary scattering is significant [1, 76, 77].

The results in Fig. 8 only considered the effect of phonon scattering. Such an analysis will relevant for the case of undoped, or charge transfer thermoelectric devices using methods as described in Refs [66, 73, 75]. As we have seen in Fig. 4, however, the effect of impurity scattering dominates the amplitude of the thermoelectric coefficients at the high impurity concentrations of $n_0=10^{19}/cm^3$ where the peak of the power factor appears. As discussed in Ref. [38], the impurity scattering limited conductivity increases with temperature because the faster moving carriers are deflected less. Of course the screening length $L_D$ (Eq. 22) increases with $\sqrt{T}$, which tends to increase scattering, but the former process dominates. Although phonon scattering tends to decrease the conductivity as the temperature raises, at $n_0=10^{19}/cm^3$ impurity scattering dominates both the amplitude and the temperature dependence, and the overall conductivity increases. The Seebeck coefficient on the other hand is scattering independent at first order, and its temperature variation is very similar to what is shown in Fig. 8b. Therefore, although the magnitude of the power factor is lower when impurities are considered, the decreasing trend in the power factor with temperature observed for the phonon-limited result, will not hold in the impurity scattering dominated case. Rather, the power factor will increase as the temperature increases because, both: i) the impurity dominated conductivity and ii) the Seebeck coefficient, both increase with temperature and will determine the power factor temperature behavior.

Figure 9 illustrates this behavior by plotting the power factor of the n-type, [111], $D$=12nm NW, for carrier concentrations $n=10^{19}/cm^3$, as a function of temperature. Figure 9a shows the phonon-limited result which has a monotonic decrease with temperature increase as in Fig. 8c. Indeed, this behavior is in agreement with recent experimental observations for gated, undoped thermoelectric devices [66]. Figure 9b shows the phonon plus impurity limited result which has the opposite trend. It demonstrates that the temperature dependence of the impurity scattering-limited conductivity dominates the temperature dependence of the power factor, rather than the temperature dependence of the phonon-limited conductivity. This behavior is also in agreement with experimental observations for the power factor in devices with heavily doped channels [1]. We mention that the magnitude of the impurity dominated result is much lower than the phonon-limited result. Since impurity scattering has such strong detrimental effect on the conductivity, utilizing methods to achieve high carrier concentrations other than direct doping would be beneficial [66, 73, 74, 75].

B. **n-type NWs: The *ZT* figure of merit**

Recent reports on thermal conductivity measurements have shown that the thermal conductance of Si NWs with diameters scaled down to $D=15$nm or $D=20$nm can be as low as $k=1$-$2$W/mK [1, 78, 79], two orders of magnitude lower than its bulk material value. As a result, the room temperature thermoelectric figure of merit (*ZT*) of silicon NWs was measured to be $ZT_{NW}\sim0.5$, greatly enhanced from the bulk Si value $ZT_{bulk}\sim0.01$. Using $k_l=2$W/mK for the lattice thermal conductivity, we estimate the expected *ZT* using the calculated power factors and $k_e$ for the NWs considered in this study. We consider separately both, the effects of phonon, SRS and impurity scattering for concentrations $10^{19}$/cm$^3$. The *ZT* values we report are just estimates since $k_l$ can be even lower, as well as orientation dependent [80, 81], in which cases *ZT* can be different and even higher.

Figure 10 shows the results for the *ZT* figure of merit versus diameter for the n-type NWs in [100] (blue), [110] (red) and [111] (green) NW orientations. Figure 10a shows phonon-limited results (dashed lines) and phonon plus SRS limited results (solid lines). The trends in all cases follow the power factor trends described in the previous sections. *ZT* values close to unity can be achieved for the larger NW diameters. For the smaller diameters, the *ZT* values are lower, even below $ZT\sim0.5$. Figure 10b shows phonon plus SRS plus impurity scattering limited results (solid lines). The phonon-limited results (dashed lines – same as in Fig. 10a) are also shown for comparison purposes. Again the influence of impurity scattering at such high concentrations causes a large degradation in the *ZT*, bringing it at values close to $ZT\sim0.2$, a factor of ~4X lower than the phonon-limited result.

### C. p-type NWs: The *ZT* figure of merit

The corresponding *ZT* results for p-type NWs are shown in Fig. 11. For p-type NWs the *ZT* values are somewhat lower than for the n-type NWs. The phonon-limited *ZT* in Fig. 11a (dashed lines) picks up for smaller diameters for the [111] and [110] NWs and reaches around unity, but at larger diameters *ZT* drops to ~0.2. SRS (solid lines in Fig. 11a) causes the *ZT* at lower diameters to reduce to *ZT*~0.5. The trends again follow the power factor trends described in Fig. 7. Finally, the influence of impurity scattering for concentrations of $p_0=10^{19}/cm^3$ in Fig. 11b (solid lines) causes a large degradation in the *ZT*, bringing it at values close to *ZT*~0.1, a factor of ~4X lower than the phonon-limited result (dashed lines).

### D. Assumptions and design optimization

The value $k_l=2$W/mK used for the calculation of *ZT* is measured for *D*=15nm Si NWs [79]. This might be even smaller for smaller NW diameters or even orientation dependent [80, 81]. In addition, the SRS strength is subject to the parameters used for the autocorrelation length $L_C$ and the $\Delta_{rms}$ value. These can be varying depending on the technology process, and be subject to the properties of the different confining surfaces [32]. The phonon scattering strength depends on the magnitude of the electron-phonon interaction used. It is suggested that the $D_{ADP}$ in NWs is as high as $D_{APD}=14.7$eV [46] instead of $D_{APD}=9.5$eV used in our calculations. Additionally, the effect of phonon confinement is not taken into account, although studies have shown that it can only result in 10~20% correction in the conductivity [43, 44, 45, 60, 62]. Depending on all these assumptions, the power factor and *ZT* could potentially change. Nevertheless, the magnitude of the results of our calculation is in agreement with other reports, both theoretical [16, 82] and experimental [1, 2, 6, 66]. More importantly, the topological trends we describe, and their importance on $\sigma$ over *S* will still hold. To that effect, the benefits of scaling the diameter in p-type [111] and [110] NWs, point toward design directions for nanoscale devices with enhanced performance.

The assumptions mentioned above affect our results mostly quantitatively. One assumption that might have qualitative influence is the fact that we have ignored the effect of NW surface relaxation and reconstruction. The electronic structure might be altered upon reconstruction, especially for the narrower $D$=3nm NWs. It is useful, therefore, to provide an estimate of how this effect can influence our results for these narrow NWs. Theoretical reports that discuss the effect of passivation and relaxation of NW surfaces using *ab-initio* methods [83, 84, 85, 86] conclude that the physical structure is still diamond-like, and does not change in any noticeable way even for ultra-thin NWs down to 1nm in diameter [84]. This is also evident from SEM and TEM images of the cross sections of NWs down to very small diameters even $D$=3nm in several instances in the literature [87, 88, 89, 90, 91]. These images show that the structure is still diamond-like, without any observable deformations. The theoretical studies show that the Si-Si surface bond lengths change only by <1.5%, whereas in the cores of the NWs the bond variation is negligible, (<0.1%) [83, 85, 92]. The bandgaps, on the other hand, change upon passivation and are strongly influenced by the choice of passivation agents (some of the common ones are -H, -OH, and –NH$_2$, and SiO$_2$) [86].

With regard to the effective masses of the subbands, which have the main influence on the electronic properties we investigate, Vo *et* al. [83] using *ab-initio* methods have calculated that upon hydrogen passivation and reconstruction, the electron and hole masses of Si NWs of $D$=3nm change as follows. For electrons: [100] ~15% increase, [110] ~7% decrease, [111] ~15% increase. For holes: [100] ~30% decrease, [110] no change, [111] no change. The p-type [100] NW is the one affected the most. As described by Buin *et* al. in Ref. [43], the mobility of single subband 1D narrow NWs follows the relation $\mu \propto m_{eff}^{-3/2}$. A 30% reduction in the effective mass will roughly result in ~70% increase in the mobility and accordingly the conductivity. Although this is a large deviation, the conductivity we report for this NW is much lower than that of the other NWs, anyway, and therefore our anisotropy conclusions will not be altered. In the case of electrons in [100] and [111] NWs the NW conductivity might be ~30% lower than what our calculations show, once relaxation is considered. Again, for this case, our results show that the conductivity for these NWs is lower than that of the [110] NW, and therefore the conclusion with regards to anisotropy are still valid. Indeed, in another work Vo *et* al. [16] reported conductivity calculations for *D*=1.1nm relaxed NWs in different transport orientations. Although a different transport model and different diameters were used, the conductivity trends they present are very similar to the trends we present in our work (but for the *D*=3nm NWs). Regarding the n-type [110] *D*=1nm NWs, Liang *et* al. [93] have also shown that reconstruction effects are minimal on the NW's ballistic performance, precisely because of the small variation of the effective mass upon reconstruction, as also calculated by Vo *et* al in Ref. [83]. We, therefore, believe that reconstruction might somewhat affect the magnitude of the conductivity we calculate in some cases, but noticeably only for the *D*=3nm p-type [100] NW. The effect, however, does not seem to be strong enough to alter the anisotropy behavior we describe.

Our results show that the influence of any design parameters, such as diameter, orientation, or even carrier type, is much stronger on the conductivity of NWs, rather than the Seebeck coefficient, which does not change significantly. The higher power factor and *ZT* appear in NWs with the highest conductivity, rather than the highest Seebeck coefficient. We note that when considering purely ballistic transport the results indicate the reverse, namely that the NWs with the largest Seebeck coefficient exhibit the largest power factors and *ZT* [11, 15]. For optimal results, the electronic structure can be engineered using quantization and band engineering. One needs to keep $\sigma$ high, by utilizing light effective mass subbands, or using strain engineering to reduce the effective masses of the subbands. In nanostructures, band engineering is partially possible, especially when utilizing devices in different orientations. From the observations about p-type [111] and [110] NWs, benefits can be achieved through proper optimization studies. Ideally one should also target the increase in *S* by allowing more valleys nearby in energy, or using transport orientations with subbands of higher degeneracy, but this should not be in the expense of $\sigma$.

# VI. Conclusion

Geometry and confinement effects on the room temperature thermoelectric coefficients ($\sigma$, $S$, $\sigma S^2$, $k_e$, $ZT$) of n- and p-type silicon NWs are investigated. Different transport orientations and diameters from $D$=12nm down to $D$=3nm are considered. Atomistic electronic structures and linearized Boltzmann transport including all relevant scattering mechanisms are employed. We find that in Si, confinement effects can have an influence on the power factor at length scales below ~7nm. We find that the influence of confinement and geometry on the power factor originates mostly from changes in the conductivity which is strongly affected, rather than the Seebeck coefficient which is weakly affected. These trends are preserved over a range of different temperatures.

In general, enhanced scattering (phonon and especially SRS) at these diameter scales strongly degrades the power factor of the device. In special cases such as p-type [110] and [111] oriented NWs, however, confinement largely improves the channel's conductivity without significantly reducing the Seebeck coefficient. Improvements in the power factor of ~2-3X can as a result be achieved. Benefits to the power factor can therefore be achieved though bandstructure modifications that influence the conductivity and not through improvements in the Seebeck coefficient.

For n-type NWs, at $D$=3nm, the [110] NW has slightly higher performance than NWs in other orientations because its subbands acquire lighter effective mass under diameter scaling, whereas at $D$=12nm the [100] NW performs better. Distinctly, the conductivity and power factor of p-type [111] and [110] NWs improves as the diameter scales below $D$=7nm. We estimate that $ZT$ values close to unity can be achieved, in agreement with experimental observations. We finally show that the maximum power factor is achieved at carrier concentrations of $10^{19}$/cm$^3$ as in the case of the bulk material. Impurity scattering strongly degrades the power factor at such concentrations. Large improvements can be achieved, however, if other possible ways to achieve high carrier concentrations other than direct doping are utilized.

# Acknowledgements

This work was supported by the Austrian Climate and Energy Fund, contract No. 825467.

Table I:

|  | ADP | IVS | SRS | Impurity |
|---|---|---|---|---|
| 1) $\Delta_{rms}=0$, $n_0=0$ | 49 | 51 | - | - |
| 2) $\Delta_{rms}^1$, $n_0=0$ | 17 | 18 | 65 | - |
| 3) $\Delta_{rms}^2$, $n_0=0$ | 6 | 6 | 88 | - |
| 4) $\Delta_{rms}^1$, $n_0^1$ | 6 | 6 | 23 | 65 |
| 5) $\Delta_{rms}^2$, $n_0^1$ | 4 | 4 | 54 | 38 |
| 6) $\Delta_{rms}^1$, $n_0^2$ | 4 | 4 | 14 | 78 |
| 7) $\Delta_{rms}^2$, $n_0^2$ | 3 | 3 | 38 | 56 |

Table I caption:

The percentage of the separate contributions of the individual scattering mechanisms to the total relaxation rate for the n-type, [100], $D=3$nm NW. The different scattering mechanisms are shown column-wise. Different combinations of scattering conditions are shown row-wise. Phonon scattering is always included, whereas SRS and impurity scattering are gradually added. In the first row only phonon scattering is considered. $\Delta_{rms}^1$ corresponds to $\Delta_{rms}=0.24$nm, and $\Delta_{rms}^2$ to $\Delta_{rms}=0.48$nm (the one used throughout the manuscript). $n_0^1$ corresponds to impurity concentration $n_0=10^{18}/cm^3$, and $n_0^2$ to impurity concentration $n_0=10^{19}/cm^3$ (the one used throughout the manuscript).

Figure 1:

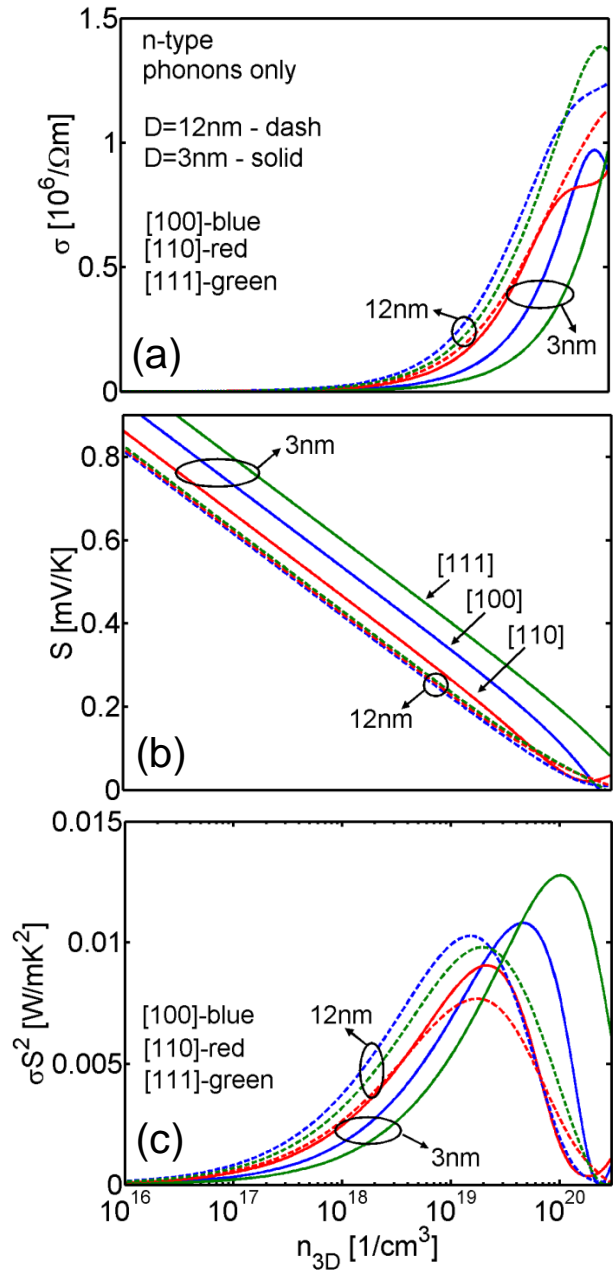

Figure 1 caption:

Thermoelectric coefficients for n-type NWs in [100] (blue), [110] (red) and [111] (green) transport orientations for diameters $D$=3nm (solid lines) and $D$=12nm (dashed lines) versus carrier concentration. (a) The electrical conductivity. (b) The Seebeck coefficient. (c) The power factor. Only phonon scattering (ADP and IVS) are considered.

Figure 2:

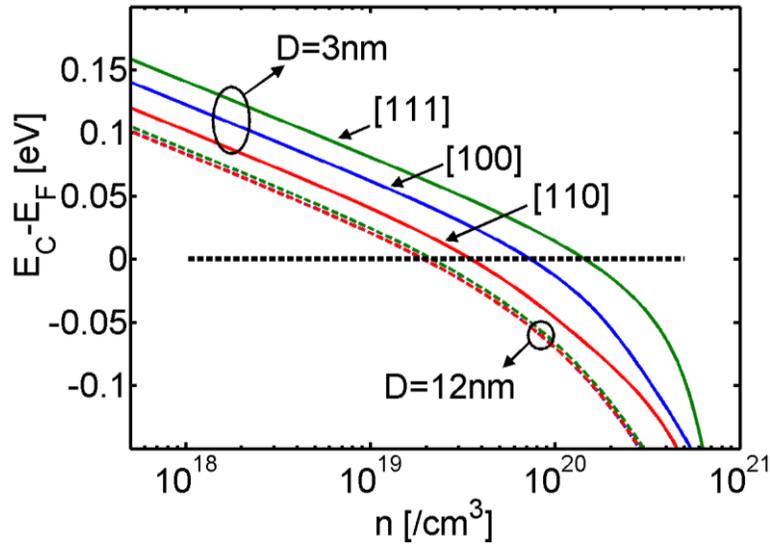

Figure 2 caption:

$E_C - E_F$ for n-type NWs of *D*=12nm (dashed lines) and *D*=3nm (solid lines) versus the carrier concentration. [100] (blue), [110] (red) and [111] (green) transport orientations are shown.

Figure 3:

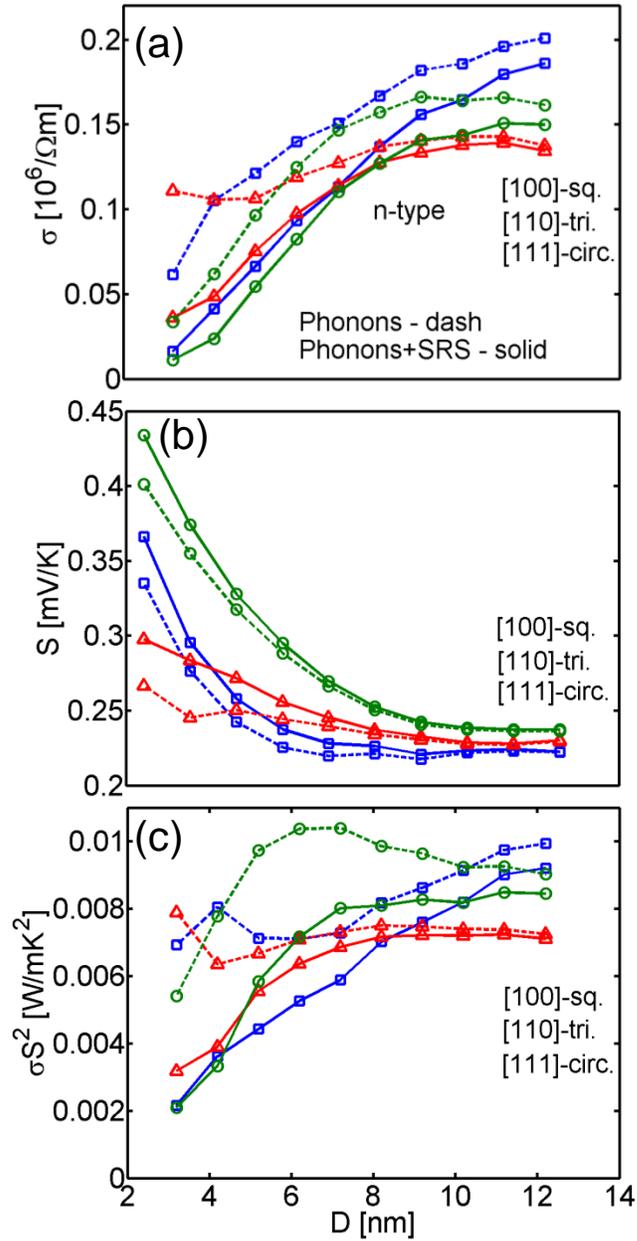

Figure 3 caption:

Thermoelectric coefficients for n-type NWs in [100] (square-blue), [110] (triangle-red), and [111] (circle-green) transport orientations versus the NW diameter. The carrier concentration is $n=10^{19}/cm^3$. (a) The electrical conductivity. (b) The Seebeck coefficient. (c) The power factor. Dashed lines: Only phonon scattering (ADP and IVS) is considered. Solid lines: Phonon scattering and SRS are considered.

Figure 4:

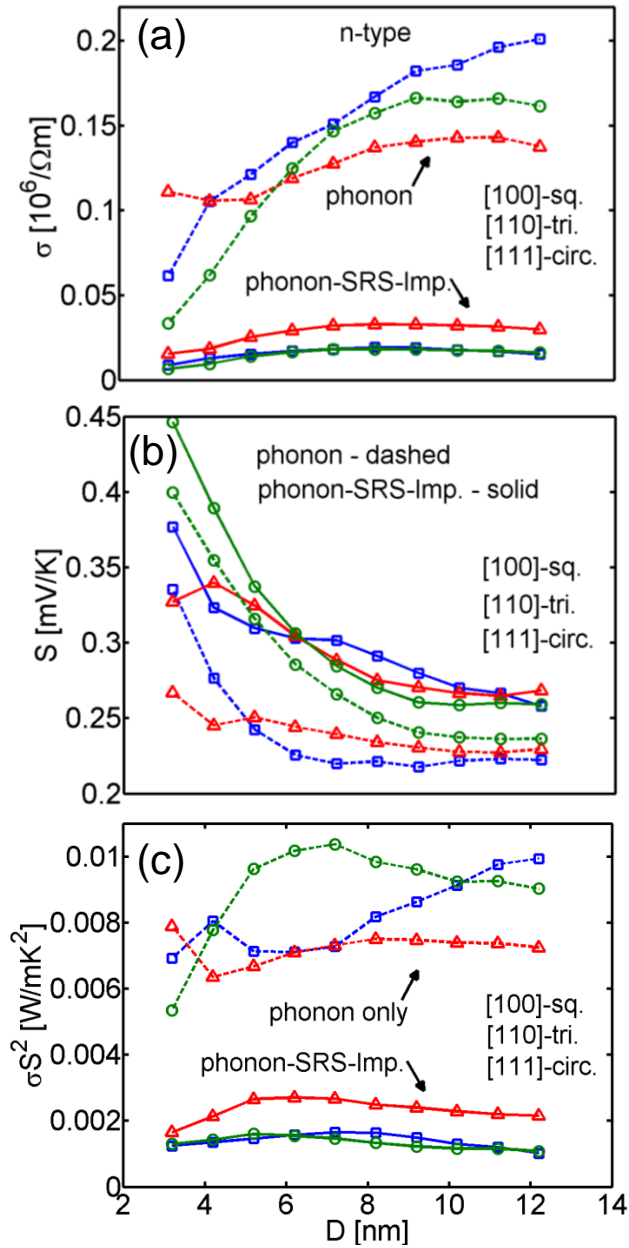

Figure 4 caption:

Thermoelectric coefficients for n-type NWs in [100] (square-blue), [110] (triangle-red), and [111] (circle-green) transport orientations versus the NW diameter. (a) The electrical conductivity. (b) The Seebeck coefficient. (c) The power factor. Solid lines: Phonon scattering, SRS and impurity scattering are considered. The impurity concentration is $n_0=10^{19}/cm^3$. Dashed lines: Only phonon scattering (ADP and IVS) is considered.

Figure 5:

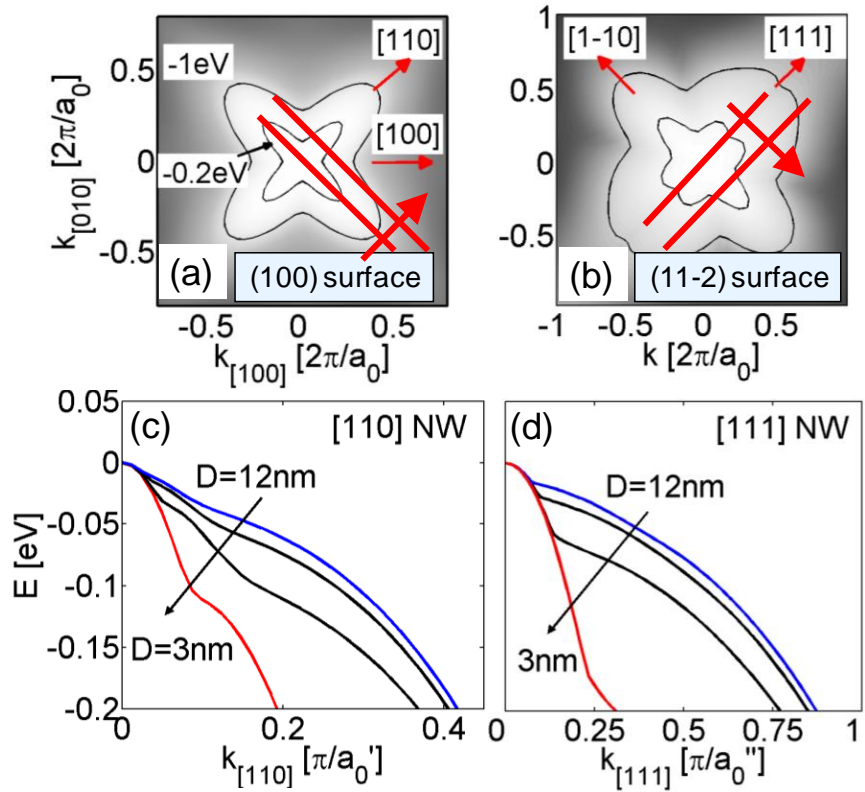

Figure 5 caption:

Bandstructure variations with confinement. (a) The (100) heavy-hole band energy surface of bulk Si. (b) The (11-2) heavy-hole band energy surface of bulk Si. Energy contours at -0.2eV and -1eV are shown. (c) The subband envelopes (higher subbands) for [110] p-type NWs as the diameter scales. (d) The subband envelopes for [111] p-type NWs as the diameter scales. Envelopes for the $D$=12, 9, 6, and 3nm are shown. $a_0$, $a_0$', and $a_0$'' are the unit cell lengths.

Figure 6:

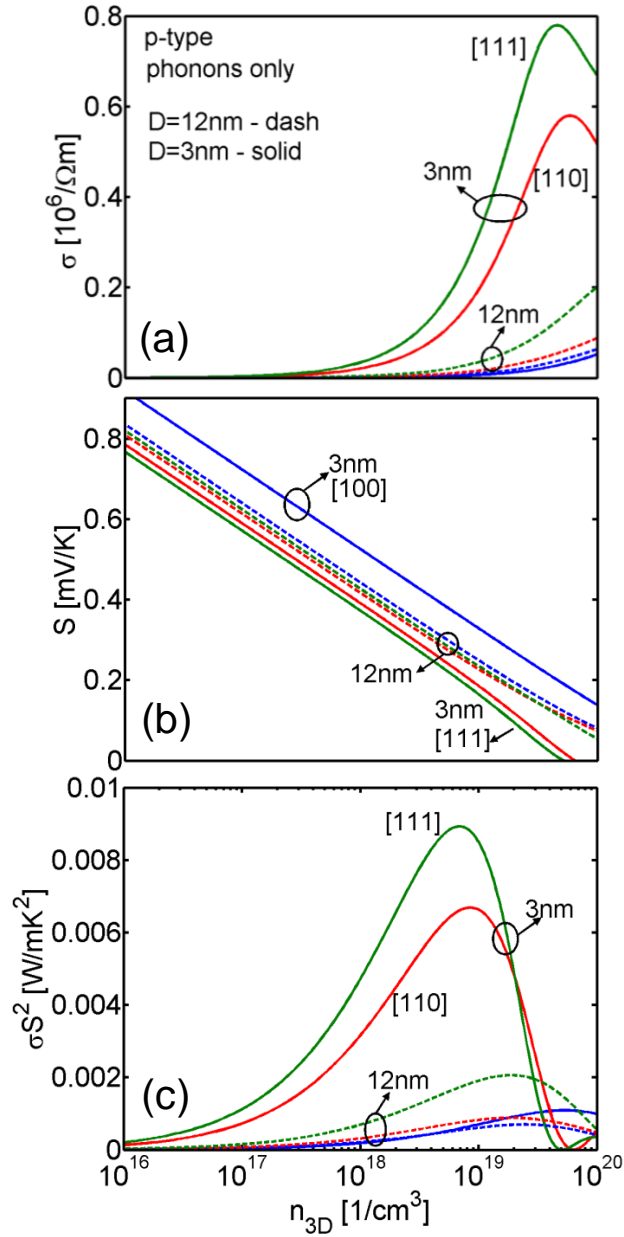

Figure 6 caption:

Thermoelectric coefficients for p-type NWs in [100] (blue), [110] (red) and [111] (green) transport orientations for diameters $D$=3nm (solid) and $D$=12nm (dashed) versus the carrier concentration. (a) The electrical conductivity. (b) The Seebeck coefficient. (c) The power factor. Only phonon scattering (ADP and ODP) are considered.

Figure 7:

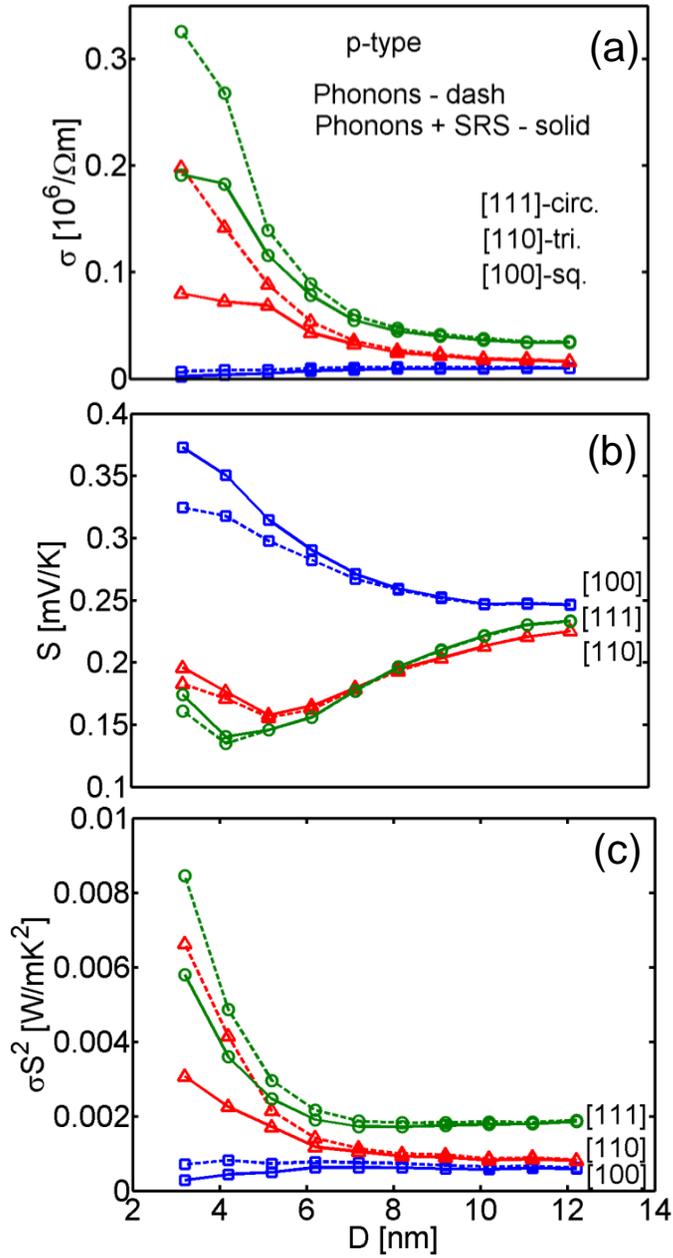

Figure 7 caption:

Thermoelectric coefficients for p-type NWs in [100] (square-blue), [110] (triangle-red) and [111] (circle-green) transport orientations versus the NW diameter. The carrier concentration is $p=10^{19}/cm^3$. (a) The electrical conductivity. (b) The Seebeck coefficient. (c) The power factor. Solid lines: Phonon scattering (ADP and ODP) and SRS are considered. Dashed lines: Only phonon scattering is considered.

Figure 8:

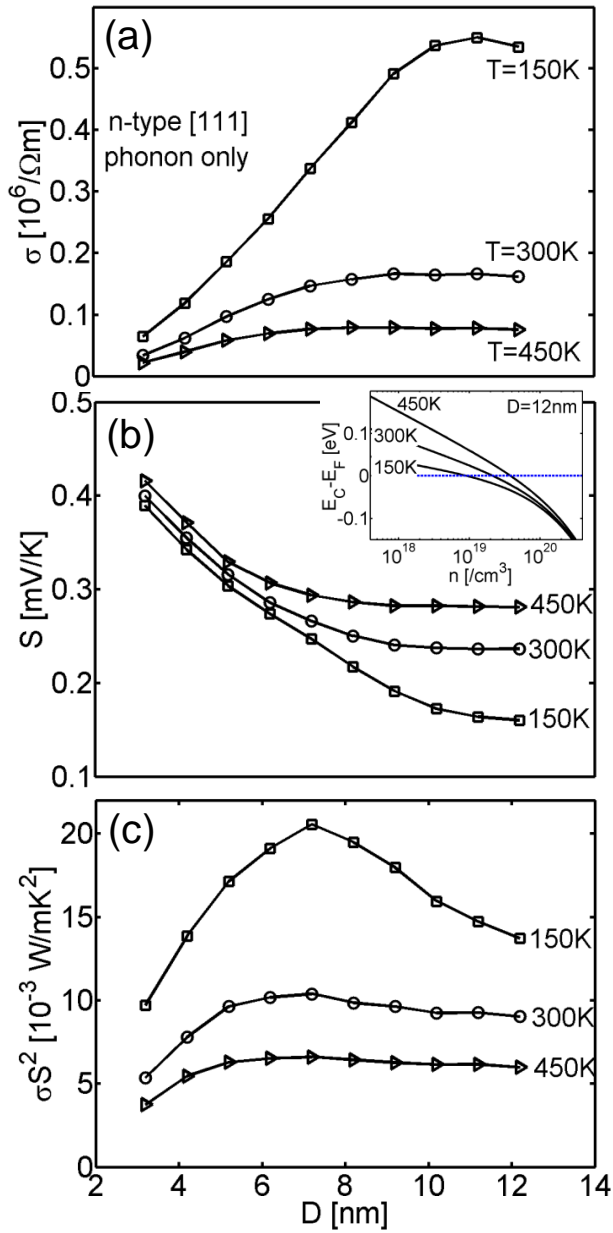

## Figure 8 caption:

Phonon-limited thermoelectric coefficients for n-type [111] NWs under different temperatures versus the NW diameter. The carrier concentration is $n=10^{19}/cm^3$. (a) The electrical conductivity. (b) The Seebeck coefficient. (c) The power factor. $T=150K$, 300K, and 450K are considered.

Figure 9:

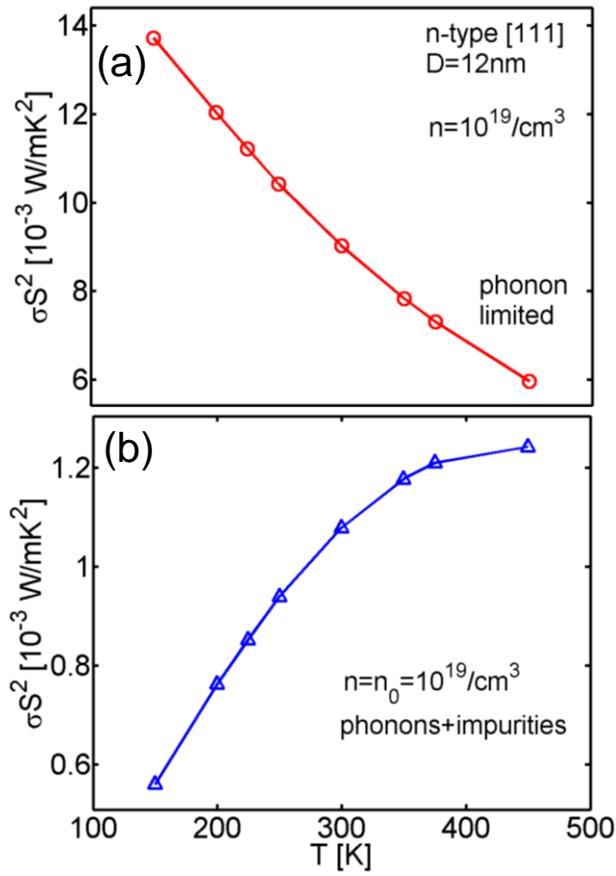

Figure 9 caption:

The power factor of the n-type [111] $D$=12nm NW versus temperature. The carrier concentration is $n=10^{19}$/cm$^3$. (a) Phonon limited results. (b) Phonon plus impurity scattering limited results, with impurity concentration $n_0=10^{19}$/cm$^3$.

Figure 10:

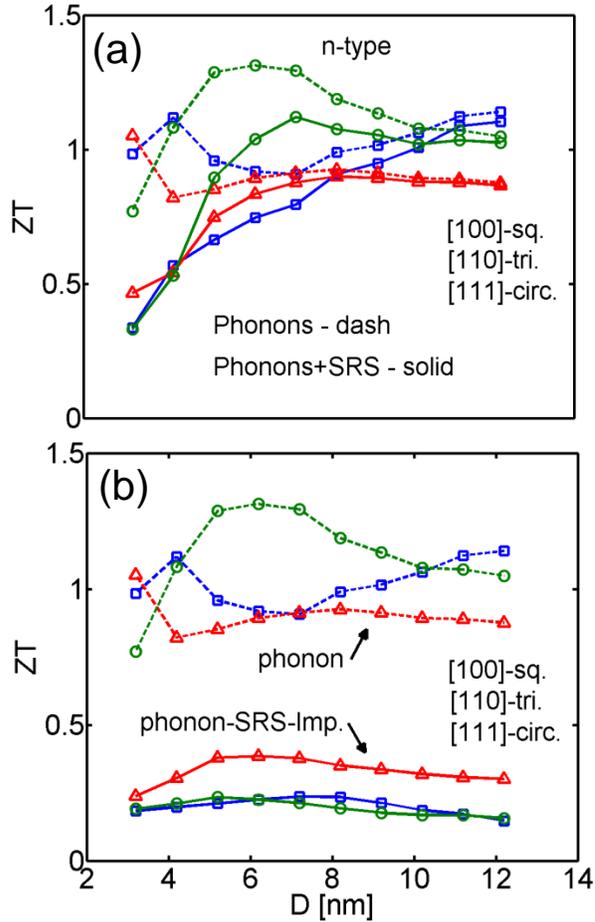

Figure 10 caption:

The *ZT* figure of merit versus diameter for n-type NWs in [100] (square-blue), [110] (triangle-red) and [111] (circle-green) transport orientations. The carrier concentration is $n=10^{19}/cm^3$. (a) Dashed lines: Only phonon scattering (ADP and IVS) is considered. Solid lines: Phonon scattering and SRS are considered. (b) Dashed lines: Only phonon scattering (ADP and IVS) is considered (same as in (a)). Solid lines: Phonon scattering, SRS and impurity scattering are considered. The impurity concentration is $n_0=10^{19}/cm^3$.

Figure 11:

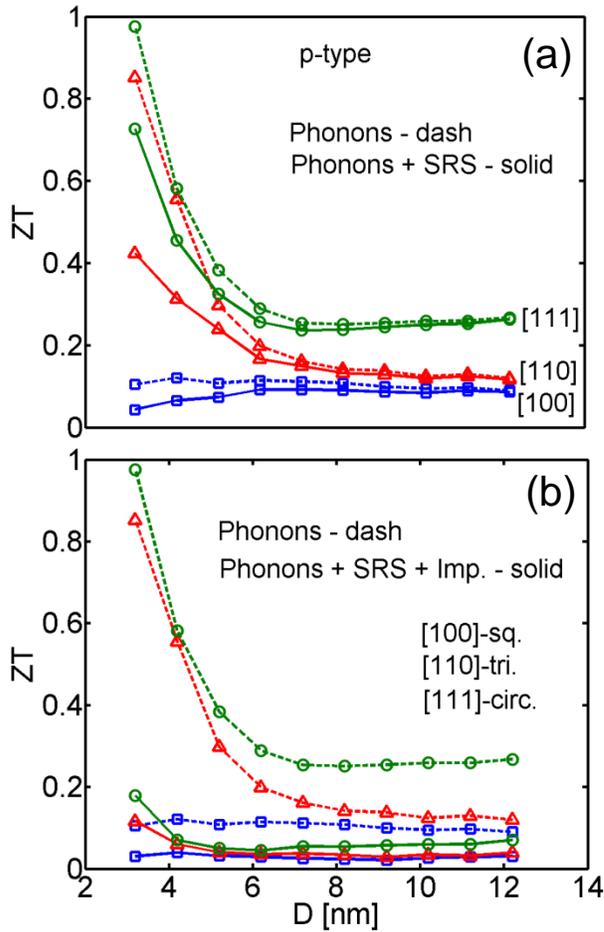

## Figure 11 caption:

The *ZT* figure of merit versus diameter for p-type NWs in [100] (square-blue), [110] (triangle-red) and [111] (circle-green) transport orientations. The carrier concentration is $p=10^{19}/cm^3$. (a) Dashed lines: Only phonon scattering (ADP and ODP) is considered. Solid lines: Phonon scattering and SRS are considered. (b) Dashed lines: Only phonon scattering (ADP and ODP) is considered (same as in (a)). Solid lines: Phonon scattering, SRS and impurity scattering are considered. The impurity concentration is $p_0=10^{19}/cm^3$.